\documentclass[
    affil-it, 
    auth-lg, 
    british, 
    contents, 
    custom-packages={scalerel}, 
    references={bigone}, 
]{lib/preprint}

\begin{document}
    
    \date{\today}
    \email{branislav.jurco@gmail.com,t.macrelli@surrey.ac.uk,c.saemann@hw.ac.uk,m.wolf@surrey.ac.uk}
    \preprint{DMUS--MP--19/11,EMPG--19--26}
    \title{Loop Amplitudes and Quantum Homotopy Algebras}
    \author[a]{Branislav Jur{\v c}o}
    \author[b]{Tommaso Macrelli}
    \author[c]{Christian S{\"a}mann}
    \author[b]{Martin Wolf}
    \affil[a]{Charles University Prague\\ Faculty of Mathematics and Physics, Mathematical Institute\\ Prague 186 75, Czech Republic}
    \affil[b]{Department of Mathematics, University of Surrey\\ Guildford GU2 7XH, United Kingdom}
    \affil[c]{Maxwell Institute for Mathematical Sciences\\ Department of Mathematics, Heriot--Watt University\\ Edinburgh EH14 4AS, United Kingdom}
    \abstract{We derive a recursion relation for loop-level scattering amplitudes of Lagrangian field theories that generalises the tree-level Berends--Giele recursion relation in Yang--Mills theory. The origin of this recursion relation is the homological perturbation lemma, which allows us to compute scattering amplitudes from minimal models of quantum homotopy algebras in a recursive way. As an application of our techniques, we give an alternative proof of the relation between non-planar and planar colour-stripped scattering amplitudes.}
    \acknowledgements{We thank Carlos Mafra, Pierpaolo Mastrolia, James Stasheff, and Alessandro Torrielli for fruitful discussions. B.J.~was supported by GA{\v C}R Grant EXPRO 19-28268X. T.M.~was partially supported by the EPSRC grant EP/N509772.}
    \datamanagement{No additional research data beyond the data presented and cited in this work are needed to validate the research findings in this work.} 
    
    \begin{body}    
        
        \section{Introduction}
        
        Batalin--Vilkovisky (BV) quantisation~\cite{Batalin:1981jr,Schwarz:1992nx} does not only help in gauge fixing and quantising complicated quantum field theories, but it also provides an important link between classical and quantum field theories and homotopy algebras even for theories without gauge symmetries. At the classical level, the BV formalism associates to every Lagrangian field theory an $L_\infty$-algebra which captures both the kinematics and the dynamics of the field theory~\cite{Costello:2007ei,Costello:2011np,Hohm:2017pnh,Jurco:2018sby,Jurco:2019bvp}. 
        
        The action of the classical field theory translates to the homotopy Maurer--Cartan action of its $L_\infty$-algebra, having the same set of fields, symmetries, equations of motions and Noether currents. Physically equivalent classical field theories have quasi-isomorphic $L_\infty$-algebras, which is the appropriate notion of equivalence from a mathematical perspective.
        
        The tree-level scattering amplitudes of a quantum field theory are encoded in the minimal models (i.e.~smallest quasi-isomorphic forms) of its $L_\infty$-algebra. Recently, it was shown that the quasi-isomorphism between both induces recursion relations for these amplitudes~\cite{Macrelli:2019afx} (see also~\cite{Arvanitakis:2019ald} for related discussions of the S-matrix in the $L_\infty$-language, \cite{Lopez-Arcos:2019hvg} for the tree-level perturbiner expansion, and~\cite{Nutzi:2018vkl,Reiterer:2019dys} for an $L_\infty$-interpretation of tree-level on-shell recursion relations). In the context of Yang--Mills (YM) theory, this recursion relation is known as the Berends--Giele recursion relation~\cite{Berends:1987me}. 
        
        In this article, we generalise the results of~\cite{Macrelli:2019afx} to loop-level scattering amplitudes. We describe our theory by a quantum homotopy algebra, and computing its minimal model via the homological perturbation lemma~\cite{gugenheim1991perturbation,Crainic:0403266}, we find a recursion relation for the loop-level scattering amplitudes. Instead of using the $L_\infty$-algebras produced by the BV-formalism directly, we use underlying unique $A_\infty$-algebras, which simplifies our discussion significantly. We apply our formalism to both scalar field theory and YM theory, and use it to analyse the one-loop $n$-gluon scattering amplitudes of~\cite{Bern:1990ux,Bern:1994zx}.
        
        \section{Scalar field theory}
        
        \subsection{Homotopy algebra}
        
        As a transparent example of our formalism, we first consider the action of a scalar field $\varphi$ with cubic and quartic interaction on four-dimensional Minkowski space $\IR^{1,3}$ with metric $\eta$,
        \begin{equation}\label{eq:phi4-action}
            S_{\rm scalar}\ \coloneqq\ -\int\rmd^{4}x~\Big\{\tfrac12\varphi\Box\varphi+\tfrac{\kappa}{3!}\varphi^3+\tfrac{\lambda}{4!}\varphi^4\Big\}~,
        \end{equation}
        where $\Box\coloneqq\eta^{\mu\nu}\partial_\mu\partial_\nu$ and $\kappa,\lambda\in\IR$. The BV formalism assigns to every classical action an $L_\infty$-algebra, cf.~\cite{Jurco:2018sby,Jurco:2019bvp} and references therein. It is possible and convenient to generalise this picture to another class of homotopy algebras known as $A_\infty$-algebras. These give rise to $L_\infty$-algebras just as the commutator on a matrix algebra induces a Lie algebra structure, and they will be useful in discussing the planar limit in Section~\ref{sec:colour}. We shall make further comments in Section~\ref{sec:conclusions}, where we explain that this generalisation is essentially unique. Importantly, the action~\eqref{eq:phi4-action} can now be identified with the homotopy Maurer--Cartan action of an $A_\infty$-algebra $\fra\cong\fra_1\oplus\fra_2$, cf.~\cite{Kajiura:2001ng,Kajiura:0306332},
        \begin{equation}\label{eq:homotopy_MC_action}
            S_{\rm hMC}\ \coloneqq\ \sum_{i=1}^\infty\frac{1}{i+1}\inner{\varphi}{\sfm_i(\varphi,\ldots,\varphi)}~,
        \end{equation}
        where the fields $\varphi$ take values in the vector space $\fra_1$ and the multi-linear maps $\sfm_i:\fra_1\times\cdots\times\fra_1\to\fra_2$ are called higher products. The vector space $\fra_2$, containing the antifields $\varphi^+$, is isomorphic to $\fra_1[-1]$ where the square brackets indicates the ghost number, and the inner product $\inner{-}{-}:\fra\times\fra\to\IR$ reads as 
        \begin{equation}
            \inner{\varphi}{\varphi^+}\ \coloneqq\ \int\rmd^{4}x~\varphi(x)\varphi^+(x)~.
        \end{equation}
        An obvious choice for the $\sfm_i$ is
        \begin{equation}\label{eq:higher_products}
            \begin{gathered}
                \sfm_1(\varphi_1)\ \coloneqq\ -\Box\varphi_1~,~~
                \sfm_2(\varphi_1,\varphi_2)\ \coloneqq\ -\tfrac12\kappa\varphi_1\varphi_2~,\\
                \sfm_3(\varphi_1,\varphi_2,\varphi_3)\ \coloneqq\ -\tfrac1{3!}\lambda \varphi_1\varphi_2\varphi_3
            \end{gathered}
        \end{equation}
        with all other higher products vanishing.
        
        To rigorously define the field space $\fra_1$ in $\fra$, we can follow~\cite{Macrelli:2019afx}. Using intuition from scattering theory, we decompose $\fra_1$ into free quantum fields and interacting or propagating ones. The former are elements of $\scC^\infty(\IR^{1,3})$ in the kernel of $\sfm_1$ with compact support on each Cauchy surface and the latter are given by the Schwartz functions $\scS(\IR^{1,3})$. This decomposition requires regularising the products~\eqref{eq:higher_products} as done in~\cite{Macrelli:2019afx} but we shall suppress these issues in the following.
        
        It is now helpful to switch to the coalgebra picture of $\fra$ which means considering the tensor algebra
        \begin{equation}
            \sfT^\bullet(\fra)\ \coloneqq\ \bigoplus_{k=0}^\infty\sfT^k(\fra)\ =\ \IR\,\oplus\,\fra\,\oplus\,(\fra\otimes\fra)\oplus\cdots,
        \end{equation}
        and extending the higher products $\sfm_i$ as coderivations $\sfM_i$ from $\fra$ to $\sfT^\bullet(\fra)$. For instance, for $\varphi_{1,\ldots,4}\in\fra_1$ we set
        \begin{equation}\label{eq:extendedProducts}
            \sfM_3(\varphi_1\otimes\cdots\otimes\varphi_4)\ \coloneqq\ \sfm_3(\varphi_1,\varphi_2,\varphi_3)\otimes\varphi_4+\varphi_1\otimes\sfm_3(\varphi_2,\varphi_3,\varphi_4)
        \end{equation}
        and $\sfM_1(\IR)=0$, $\sfM_2(\varphi_1)=0$, etc. These coderivations combine into a linear map $\sfD:\sfT^\bullet(\fra)\to\sfT^\bullet(\fra)$,
        \begin{equation}
            \sfD\ \coloneqq\ \sfM_1+\sfM_2+\sfM_3~,
        \end{equation}
        which is a codifferential. In fact, an $A_\infty$-algebra can be defined to be a $\IZ$-graded vector space with a codifferential on its tensor algebra.
        
        \subsection{Tree-level scattering amplitudes}
        
        For every $A_\infty$-algebra $\fra$, the product $\sfm_1$ is a differential on $\fra$. Consequently, one can study the cohomology $H^\bullet_{\sfm_1}(\fra)$ with respect to $\sfm_1$, and, for instance, $H^1_{\sfm_1}(\fra)$ contains all free on-shell fields. This cohomology extends to an $A_\infty$-algebra $(\fra^\circ\coloneqq H^\bullet_{\sfm_1}(\fra),\sfm_i^\circ)$ with $\sfm_1^\circ=0$, called the minimal model, which encodes the $n$-point tree-level scattering amplitudes, cf.~\cite{Macrelli:2019afx} (see also~\cite{Kajiura:2001ng,Kajiura:0306332}),
        \begin{equation}\label{eq:treelevelScalarAmplitude}
            \begin{aligned}
                \caA_{n,0}(\varphi_1,\ldots,\varphi_n)
                \ &=\ \sum_{\sigma\in S_{n-1}}\inner{\varphi_n}{\sfm^\circ_{n-1}(\varphi_{\sigma(1)},\ldots,\varphi_{\sigma(n-1)})}\\
                \ &=\ \sum_{\sigma\in S_n/\IZ_n}\inner{\varphi_{\sigma(1)}}{\sfm^\circ_{n-1}(\varphi_{\sigma(2)},\ldots,\varphi_{\sigma(n)})}~,
            \end{aligned}
        \end{equation}
        where the $\varphi_i\in H^1_{\sfm_1}(\fra)$ are free fields.
        
        The relation between $\fra^\circ$ and $\fra$ itself is best depicted by the diagram
        \begin{subequations}\label{eq:contractingHomotopy}
            \begin{equation}\label{eq:initial_contraction}
                \begin{tikzcd}
                    \ar[loop,out=160,in=200,distance=20,"\sfh" left] (\fra,\sfm_1)\arrow[r,twoheadrightarrow,shift left]{}{\sfp} & (\fra^\circ,0) \arrow[l,hookrightarrow,shift left]{}{\sfe}~,
                \end{tikzcd}
            \end{equation}
            where $\sfp$ is the obvious projection, $\sfe$ is a choice of embedding, and $\sfh$ is the propagator, i.e.~the inverse of $\sfm_1$ on the Schwartz functions $\scS(\IR^{1,3})$ trivially continued to $\fra_2$ such that its kernel is the cokernel $\sfe(\fra_2^\circ)$ of $\sfm_1$. The maps $\sfe$ and $\sfh$ can be chosen such that 
            \begin{equation}\label{eq:contractingBasic}
                \begin{gathered}
                    1\ =\ \sfm_1\circ\sfh+\sfh\circ \sfm_1+\sfe\circ\sfp~,\\
                    \sfp\circ\sfe\ =\ 1~,\\
                    \sfp\circ\sfh\ =\ \sfh\circ\sfe\ =\ \sfh\circ\sfh\ =\ 0~,\\
                    \sfp\circ \sfm_1\ =\ \sfm_1\circ\sfe\ =\ 0~.
                \end{gathered}
            \end{equation}
        \end{subequations}
        Mathematically, this is known as an abstract Hodge--Kodaira decomposition.\footnote{In this context, the propagator $\sfh$ is a chain homotopy, a fact that is explained in~\cite{Mnev:2006ch,Mnev:2008sa,Cattaneo:0811.2045}, see also~\cite{Cattaneo:2015vsa} and references therein.} 
        
        The higher products of the minimal model $\sfm_i^\circ$ are again encoded in a codifferential $\sfD^\circ$ on $\sfT^\bullet(\fra^\circ)$. This follows from the homological perturbation lemma~\cite{gugenheim1991perturbation,Crainic:0403266}, which also gives a prescription of how to compute $\sfD^\circ$. 
        
        We can extend both $\sfp$ and $\sfe$ trivially to corresponding maps $\sfP_0$ and $\sfE_0$ between $\sfT^\bullet(\fra)$ and $\sfT^\bullet(\fra^\circ)$,
        \begin{subequations}
            \begin{equation}
                \sfP_0|_{\sfT^k(\fra)}\ \coloneqq\ \sfp^{\otimes^k}\eand
                \sfE_0|_{\sfT^k(\fra^\circ)}\ \coloneqq\ \sfe^{\otimes^k}~.
            \end{equation}
            The propagator $\sfh$ is extended to a map $\sfH_0:\sfT^\bullet(\fra)\to\sfT^\bullet(\fra)$ via the tensor trick,
            \begin{equation}\label{eq:contractingHomotopyZero}
                \begin{gathered}
                    \sfH_0|_{\sfT^k(\fra)}\ \coloneqq\ \sum_{i+j=k-1}1^{\otimes^i}\otimes\sfh\otimes(\sfe\circ\sfp)^{\otimes^j}~.
                \end{gathered}
            \end{equation}
        \end{subequations}
        Splitting $\sfD$ into the `free' part $\sfD_0\coloneqq\sfM_1$ and the `interaction' part $\sfD_\rint\coloneqq\sfM_2+\sfM_3$, we recover~\eqref{eq:contractingHomotopy} with the maps $\sfm_1$, $\sfp$, $\sfe$, and $\sfh$ replaced by $\sfM_1$, $\sfP_0$, $\sfE_0$, and $\sfH_0$.
        
        The homological perturbation lemma allows us to deform $\sfM_1$ to the codifferential $\sfD$, regarding $\sfD_\rint$ as a perturbation, which induces a codifferential $\sfD^\circ$ on $\sfT^\bullet(\fra^\circ)$,
        \begin{equation}\label{eq:hpl_relations}
            \begin{gathered}
                \sfP\ =\ \sfP_0\circ(1+\sfD_\rint\circ\sfH_0)^{-1},~~
                \sfH\ =\ \sfH_0\circ(1+\sfD_\rint\circ\sfH_0)^{-1}~,\\
                \sfE\ =\ (1+\sfH_0\circ\sfD_\rint)^{-1}\circ\sfE_0,~~
                \sfD^\circ\ =\ \sfP\circ\sfD_\rint\circ\sfE_0~.
            \end{gathered}
        \end{equation}
        We have again diagram~\eqref{eq:initial_contraction} and relations~\eqref{eq:contractingBasic} with the maps $\sfm_1$, $\sfp$, $\sfe$, and $\sfh$ replaced by $\sfD$, $\sfP$, $\sfE$, and $\sfH$. Moreover, $\sfE$ and $\sfP$ satisfy the evident relations
        \begin{equation}
            \sfP\circ\sfD\ =\ \sfD^\circ\circ\sfP\eand\sfD\circ\sfE\ =\ \sfE\circ\sfD^\circ~.
        \end{equation}
        The equations for $\sfE$ and $\sfH$ in~\eqref{eq:hpl_relations} imply
        \begin{subequations}\label{eq:treelevelE}
            \begin{eqnarray}
                \sfD^\circ\!&=&\!\sfP_0\circ\sfD_\rint\circ\sfE~,\label{eq:treelevelE1}\\
                \sfE\!&=&\!\sfE_0-\sfH_0\circ\sfD_\rint\circ\sfE~.\label{eq:treelevelE2}
            \end{eqnarray}
        \end{subequations}
        Substituting~\eqref{eq:treelevelE2} back into itself yields a recursion relation in the powers of the coupling constants since $\sfD_\rint$ adds one power of either $\kappa$ or $\lambda$. Equation~\eqref{eq:treelevelE1} then allows us to construct $\sfD^\circ=\sum_{i=2}^\infty\sfM_i^\circ$ and hence, the products $\sfm_i^\circ$ entering the amplitude~\eqref{eq:treelevelScalarAmplitude}. By construction, $\sfM_1^\circ=0$ and so $\sfm_1^\circ=0$. If we restrict the action of $\sfE$ to $\sfT^n(\fra^\circ)$ and project the result onto $\fra=\sfT^1(\fra)\subseteq\sfT^\bullet(\fra)$, we recover the tree-level $n$-point Berends--Giele current for scalar field theory (see our paper~\cite{Macrelli:2019afx} for full details).
        
        \subsection{Loop-level scattering amplitudes}
        
        The BV formalism gives a clear indication as how to generalise the above to the quantum case: the codifferential $\sfD$ of the previous section is the dual of the classical BV differential. In the quantum case, the term $-\rmi\hbar\Delta$ is added to this differential, where $\Delta$ is the usual BV Laplacian featuring in the quantum master equation~\cite{Batalin:1981jr,Schwarz:1992nx}. In the coalgebra picture, this amounts to adding $-\rmi\hbar\Delta^*$ which inserts a complete set of fields and antifields in any possible way into the tensor product, preserving the order of the original factors. For $\varphi_{1,2}\in\fra$, for example,
        \begin{equation}\label{eq:actionDualBVLScalar}
            \begin{aligned}
                \Delta^*(\varphi_1\otimes\varphi_2)\ &=\ \int\frac{\rmd^4k}{(2\pi)^4}\Big\{\psi(k)\otimes\psi^+(k)\otimes\varphi_1\otimes\varphi_2+\psi(k)\otimes\varphi_1\otimes\psi^+(k)\otimes\varphi_2+\cdots+\\
                &\kern2cm+\psi^+(k)\otimes\psi(k)\otimes\varphi_1\otimes\varphi_2+\psi^+(k)\otimes\varphi_1\otimes\psi(k)\otimes\varphi_2+\cdots\Big\}~,
            \end{aligned}
        \end{equation}
        where $\psi(k)$ is a (momentum space) basis of the field space $\fra_1$ and $\psi^+(k)$ of the antifield space $\fra_2$.
        
        To compute the loop-level scattering amplitudes, we replace the perturbation,
        \begin{equation}\label{eq:substitution}
            \sfD_\rint\ \to\ \sfD_\rint-\rmi\hbar\Delta^*~,
        \end{equation}
        in the homological perturbation lemma (see also~\cite{Pulmann:2016aa,Doubek:2017naz}). This generalises~\eqref{eq:treelevelE} to
        \begin{subequations}\label{eq:looplevelE}
            \begin{eqnarray}
                \sfD^\circ\!&=&\!\sfP_0\circ(\sfD_\rint-\rmi\hbar\Delta^*)\circ\sfE~,\label{eq:looplevelE1}\\
                \sfE\!&=&\!\sfE_0-\sfH_0\circ(\sfD_\rint-\rmi\hbar\Delta^*)\circ\sfE~.\label{eq:looplevelE2}
            \end{eqnarray}
        \end{subequations}
        Contrary to the tree-level case, $\sfP$ and $\sfE$ are no longer coalgebra morphisms but only morphisms of graded vector spaces. Importantly, the substitution~\eqref{eq:substitution} is justified for any theory whose classical BV action also satisfies the quantum master equation. This includes scalar field theory, Chern--Simons theory, and also Yang--Mills theory.
        
        As before,~\eqref{eq:looplevelE} yields a recursion relation, now in the powers of both the coupling constants and $\hbar$. The former counts the number of interaction vertices while the latter counts the number of loops.\footnote{When a classical BV action does not satisfy the quantum master equation, one first has to construct the quantum BV action which is given as a series expansion in powers of $\hbar$. In this case, the parameter $\ell$ in \protect\eqref{eq:recursionRelation} is no longer the loop expansion parameter.} The map $\sfE$ encodes all currents, and we introduce the restrictions to $j$ factors in the input and $i$ factors in the output tensor product,
        \begin{equation}
            \sfE^{i,j}\ \coloneqq\ \big(\pr_{\sfT^i(\fra)}\circ\sfE\big)\big|_{\sfT^j(\fra^\circ)}\eand
            \sfD_\rint^{i,j}\ \coloneqq\ \big(\pr_{\sfT^i(\fra)}\circ\sfD_\rint\big)\big|_{\sfT^j(\fra)}~.
        \end{equation}
        If we further restrict to currents with $\ell$ loops and $v$ vertices,~\eqref{eq:looplevelE} becomes the recursion relation
        \begin{equation}\label{eq:recursionRelation}
            \begin{aligned}
                \sfE^{i,j}_{\ell,v}\ &=\ \delta_\ell^0\delta_v^0\delta^{ij}\sfE_0|_{\sfT^i(\fra^\circ)}-\sfH_0|_{\sfT^i(\fra)}\circ\sum_{k=2}^{i+2}\sfD_\rint^{i,k}\circ\sfE^{k,j}_{\ell,v-1}+\rmi\hbar\,\sfH_0|_{\sfT^i(\fra)}\circ\Delta^*|_{\sfT^{i-2}(\fra)}\circ\sfE^{i-2,j}_{\ell-1,v}
            \end{aligned}
        \end{equation}
        for scalar field theory. Here, we put $\sfE^{i,j}_{\ell,v}=0$ for $\ell<0$ or $v<0$ and this implies that the recursion relation terminates for each finite number of $\ell$ and $v$. 
        
        Just as the currents $\sfE$, we can also decompose the higher products according to their loop order, $\sfm^\circ_i=\sum_{\ell=0}^\infty\hbar^\ell\sfm_{i,\ell}^\circ$ with $\sfm^\circ_{1,0}=0$. The $\ell$-loop scattering amplitude reads as 
        \begin{equation}\label{eq:looplevelScalarAmplitude}
            \begin{aligned}
                \caA_{n,\ell}(\varphi_1,\ldots,\varphi_n)
                \ &=\ \sum_{\sigma\in S_{n-1}}\inner{\varphi_n}{\sfm^\circ_{n-1,\ell}(\varphi_{\sigma(1)},\ldots,\varphi_{\sigma(n-1)})}\\
                \ &=\ \sum_{\sigma\in S_n/\IZ_n}\inner{\varphi_{\sigma(1)}}{\sfm^\circ_{n-1,\ell}(\varphi_{\sigma(2)},\ldots,\varphi_{\sigma(n)})}~.
            \end{aligned}
        \end{equation}
        Mathematically, $(\fra^\circ\coloneqq H^\bullet_{\sfm_1}(\fra),\sfm_{i}^\circ)$ constitutes (the minimal model of) a quantum $A_\infty$-algebra. 
        
        \section{Yang--Mills theory}
        
        \subsection{Homotopy algebra}

        Consider again four-dimensional Minkowski space $\IR^{1,3}$ with metric $\eta$. Let $\Omega^\bullet\coloneqq\Omega^\bullet(\IR^{1,3})$ be the differential forms on $\IR^{1,3}$, $\rmd$ the exterior derivative, and $\star$ the Hodge star operator with respect to $\eta$. We also set $\rmd^\dagger\coloneqq\star\rmd\star$. The BV formalism of $\sfU(N)$ YM theory has a gauge potential $A\in\Omega^1[0]\otimes\fru(N)$ with curvature $F\coloneqq\rmd A+\tfrac\kappa2[A,A]\in\Omega^2[0]\otimes\fru(N)$ and a ghost $c\in\Omega^0[1]$. The square brackets denote the ghost degree and $[-,-]$ is the Lie bracket on $\fru(N)$ with the wedge product understood. The corresponding antifields are $A^+\in\Omega^1[-1]\otimes\fru(N)$ and $c^+\in\Omega^0[-2]\otimes\fru(N)$. Letting $\nabla\coloneqq\rmd+\kappa[A,-]$ and `tr' be the matrix trace, the BV action of YM theory is~\cite{Batalin:1981jr}
        \begin{equation}
            S_{\rm YM}\ \coloneqq\ \int\tr\Big\{\tfrac12 F\wedge\star F-A^+\wedge\star\nabla c-\tfrac\kappa2c^+\wedge\star[c,c]\Big\}~.
        \end{equation}
        Importantly, this action satisfies both the classical as well as the quantum master equations.

        Gauge fixing needs the trivial pair $(b,\bar c)\in(\Omega^0[0]\oplus\Omega^0[-1])\otimes\fru(N)$ together with the antifields pair $(b^+,\bar c^+)\in(\Omega^0[-1]\oplus\Omega^0[0])\otimes\fru(N)$ entering via
        \begin{equation}\label{eq:CBVYMTP}
            S_{\rm YM,tp}\ \coloneqq\ S_{\rm YM}-\int\tr\big\{b\wedge\star\bar c^+\big\}~,
         \end{equation}
        and it is achieved by a canonical transformation 
        \begin{equation}
            S_{\rm YM, gf}[a,\tilde a^+]\ \coloneqq\ S_{\rm YM,tp}[a,\tilde a^++\tfrac{\delta\Psi}{\delta a}]
        \end{equation}
        mediated by a choice of gauge fixing fermion $\Psi$, the generating functional of the canonical transformation, of ghost degree~$-1$. Here, we collectively denote all the fields by $a$ and all the antifields by $a^+$. We take $\Psi$ to be
        \begin{equation}\label{eq:GFF}
            \Psi\ \coloneqq\ \int\tr\Big\{\bar c\wedge\star\big(\rmd^\dagger A-\tfrac\xi2 b\big)\Big\}
        \end{equation}
        with $\xi\in\IR$ which amounts to Lorenz gauge. Explicitly, upon slightly abusing notation and denoting the transformed antifields again by the same letters, we have
        \begin{equation}\label{eq:CGFYMA}
            S_{\rm YM,gf}\ =\ \int\tr\Big\{\tfrac12 F\wedge\star F-(A^++\rmd\bar c)\wedge\star\nabla c-\tfrac\kappa2c^+\wedge \star[c,c]-b\wedge\star\big(\bar c^++\rmd^\dagger A-\tfrac\xi2 b\big)\Big\}~.
        \end{equation}
        
        As often convenient in the BV formalism, we regard all fields, ghosts, trivial pairs, antifields, etc.~as forming a superfield $\sfa$ generating the vector space\footnote{On a technical note, the vector space $\fra_1$ should again be decomposed into free (i.e.~compactly supported on Cauchy surfaces) and interacting (i.e.~Schwartz type) fields as before for scalar field theory. We shall suppress this issue in the following.} $\fra_1$, cf.~\cite{Jurco:2018sby,Jurco:2019bvp}. Working with $A_\infty$-algebras amounts to working in the `color flow' formalism or using double line Feynman diagrams. This implies that the fields take values in a matrix algebra and thus, we have to extend the gauge algebra from $\fru(n)$ to $\frgl(n,\IC)$. Similarly to scalar field theory, we add an isomorphic space of additional antifields $\fra_2\cong\fra_1[-1]$. Our $A_\infty$-algebra has then the underlying vector space\footnote{The $L_\infty$-algebra underlying YM theory was explained in~\cite{Movshev:2003ib,Movshev:2004aw,Zeitlin:2007vv,Zeitlin:2007yf}, see also~\cite{Jurco:2018sby,Jurco:2019bvp,Macrelli:2019afx}. Our $A_\infty$-algebra $\fra$ arises from an $A_\infty$-algebra extension of this $L_\infty$-algebra.} $\fra\cong\fra_1\oplus\fra_2$ and is endowed with a cyclic structure defined by
        \begin{equation}
            \stretchleftright[10000]{<}{
                \begin{pmatrix}
                    c \\ A \\ b \\ \bar c^+ \\ \bar c \\ b^+ \\ A^+ \\ c^+
                \end{pmatrix},
                \begin{pmatrix}
                    \hat c \\ \hat A \\ \hat b \\ \hat{\bar c}^+ \\ \hat {\bar c} \\ \hat b^+ \\ \hat A^+ \\ \hat c^+
                \end{pmatrix}
            }{>}
            \ =\ \begin{aligned}
                &\int\tr\,\Big\{\hat c^\dagger\wedge\star c^+-\hat A^\dagger\wedge\star A^+-\\
                &\kern1cm-\hat b^\dagger\wedge\star b^++(\hat{\bar c}^+)^\dagger\wedge\star\bar c\,\,+\\
                &\kern1cm+c^\dagger\wedge\star \hat c^+-A^\dagger\wedge\star \hat A^+\,+\\
                &\kern1cm-b^\dagger\wedge\star \hat b^++(\bar c^+)^\dagger\wedge\star\hat{\bar c}\Big\}
            \end{aligned}
        \end{equation}
        for $c,A,\ldots\in\fra_1$ and $\hat c,\hat A,\ldots\in\fra_2$. Expanding all Lie brackets in the action as matrix commutators and considering all cyclic orderings of all terms with equal weight, we can directly read off the higher products which reproduce the action~\eqref{eq:CGFYMA} from the homotopy Maurer--Cartan action
        \begin{equation}\label{eq:hMC_YM}
            S_{\rm hMC}\ \coloneqq\ \sum_{i=1}^\infty\frac{1}{i+1}\inner{\sfa}{\sfm_i(\sfa,\ldots,\sfa)}~,
        \end{equation}
        where $\sfa\coloneqq c_1+A_1+\cdots +\bar c_1+c_1^+\in \fra_1$. The non-trivial ones are
        \begin{subequations}\label{eq:YM_higher_products}
            \begin{equation}
                \sfm_1
                \begin{pmatrix}
                    c_1\\ A_1\\ b_1\\ \bar c^+_1\\ \bar c_1\\ b^+_1\\ A^+_1\\ c^+_1
                \end{pmatrix}
                \ \coloneqq\
                \begin{pmatrix}
                    0\\ -\rmd c_1\\ 0\\ \rmd^\dagger\rmd c_1\\ b_1\\ -\rmd^\dagger A_1 -\xi b_1 -\bar c^+_1\\ \rmd^\dagger\rmd A_1 -\rmd b_1\\ -\rmd^\dagger (A_1^++\rmd \bar c_1)
                \end{pmatrix}~,
            \end{equation}
            and
            \begin{equation}
                \begin{gathered}
                    \sfm_2\left(
                    \begin{pmatrix}
                        c_1\\ A_1\\ b_1\\ \bar c^+_1\\ \bar c_1\\ b^+_1\\ A^+_1\\ c^+_1
                    \end{pmatrix},
                    \begin{pmatrix}
                        c_2\\ A_2\\ b_2\\ \bar c^+_2\\ \bar c_2\\ b^+_2\\ A^+_2\\ c^+_2
                    \end{pmatrix}\right)
                    \ \coloneqq\ \kappa
                    \begin{pmatrix}
                        c_1c_2\\ c_1A_2 +A_1 c_2\\ 0\\ -\rmd^\dagger (c_1A_2+A_1c_2)\\ 0\\ 0\\ -c_1(A_2^++\rmd \bar c_2)+(A_1^++\rmd \bar c_1)c_2+\rmd^\dagger(A_1\wedge A_2)+\\\hspace{2cm}+{\star(A_1\wedge {\star\rmd A_2})}-{\star(\star{\rmd A_1}\wedge A_2)}\\ c_1c_2^+-c_2^+c_1-\star(A_1\wedge \star(A_2^++\rmd \bar c_2))+\\
                        \hspace{2cm}+\star((A_1^++\rmd \bar c_1)\wedge \star A_2)
                    \end{pmatrix}~,\\
                    \sfm_3\left(
                    \begin{pmatrix}
                        c_1\\ A_1\\ b_1\\ \bar c^+_1\\ \bar c_1\\ b^+_1\\ A^+_1\\ c^+_1
                    \end{pmatrix},
                    \begin{pmatrix}
                        c_2\\ A_2\\ b_2\\ \bar c^+_2\\ \bar c_2\\ b^+_2\\ A^+_2\\ c^+_2
                    \end{pmatrix},
                    \begin{pmatrix}
                        c_3\\ A_3\\ b_3\\ \bar c^+_3\\ \bar c_3\\ b^+_3\\ A^+_3\\ c^+_3
                    \end{pmatrix}
                    \right)
                    \ \coloneqq\ \kappa^2
                    \begin{pmatrix}
                        0\\ 0\\ 0\\ 0\\ 0\\ 0\\ {\star(A_1\wedge\star(A_2\wedge A_3))}-{\star(\star(A_1\wedge A_2)\wedge A_3)}\\ 0
                    \end{pmatrix}~.
                \end{gathered}
            \end{equation}
        \end{subequations}
        Altogether, $(\fra,\sfm_i,\langle-,-\rangle)$ is a cyclic $A_\infty$-algebra and its homotopy Maurer--Cartan action~\eqref{eq:hMC_YM} reproduces the gauge-fixed BV action~\eqref{eq:CGFYMA}.
        
        As before, scattering amplitudes are encoded in the corresponding minimal model and given by formulas of the form~\eqref{eq:treelevelScalarAmplitude} and~\eqref{eq:looplevelScalarAmplitude} with $\varphi$ replaced by $\sfa$. To determine these from the homological perturbation lemma, we note that the relevant propagator $\sfh$, which also gives rise to $\sfH_0$ via~\eqref{eq:contractingHomotopyZero}, acts as 
        \begin{equation}
            \sfh
            \begin{pmatrix}
                \hat c\\ \hat A\\ \hat b\\ \hat{\bar c}^+\\ \hat{\bar c}\\ \hat b^+\\ \hat A^+\\ \hat c^+
            \end{pmatrix}
            \ \coloneqq\
            \begin{pmatrix}
                -G^F\rmd^\dagger\hat A\\ G^FP_{\rmd^\dagger\rmd}\hat A^+\\ G^F\rmd^\dagger\hat A^++G^F\rmd^\dagger\rmd\hat{\bar c}-\xi\hat{\bar c}-\hat b^+\\ \hat{\bar c}\\ -G^FP_{\rmd^\dagger}\hat c^+\\ 0\\ -G^F\rmd\hat c^+\\ 0 
            \end{pmatrix}~,
        \end{equation}
        where $P_{(-)}$ is the projection onto $\im(-)$ and $G^F$ the Feynman--Green operator for $\Box\coloneqq\rmd^\dagger\rmd+\rmd\rmd^\dagger$ on functions. We have again a diagram~\eqref{eq:initial_contraction} and maps satisfying relations~\eqref{eq:contractingBasic}. The homological perturbation lemma then yields recursion relations of a similar form as~\eqref{eq:recursionRelation} since we again have 3- and 4-point vertices.
        
        \subsection{Colour structure of scattering amplitudes}\label{sec:colour}
        
        To demonstrate the power of our formalism, we examine the colour structure of scattering amplitudes in YM theory. This is facilitated by our generalisation from the $L_\infty$-algebras from the BV formalism to $A_\infty$-algebras.
        
        Consider plane waves $A_i=a_i X_i=a_{i\mu}\,\rmd x^\mu X_i\in H^1_{\sfm_1}(\fra)$
        with $a_{i \mu}\coloneqq\varepsilon_\mu(k_i)\,\rme^{\rmi k_i\cdot x}$, where $k_i$ is the on-shell momentum, $\varepsilon(k_i)$ is the polarisation in Lorenz gauge $k_i\cdot\varepsilon(k_i)=0$, and $X_i\in\fru(N)$ is the colour part. The scattering amplitude then is
        \begin{subequations}\label{eq:computing_amplitudes}
            \begin{equation}\label{eq:amplitude_YM}
                \begin{aligned}
                    \caA_n(A_1,A_2,\ldots,A_n)
                    \ &=\ \sum_{\sigma\in S_{n-1}}\inner{A_{n}}{\sfm^\circ_{n-1}(A_{\sigma(1)},\ldots,A_{\sigma(n-1)})}\\
                    \ &=\ \sum_{\sigma\in S_n/\IZ_n}\inner{A_{\sigma(1)}}{\sfm^\circ_{n-1}(A_{\sigma(2)},\ldots,A_{\sigma(n)}}~,
                \end{aligned}
            \end{equation}
            where 
            \begin{equation}\label{eq:minimalProductYM1}
                \sfm_{i}^\circ\ =\ \big({\rm pr}_{\sfT^1(\fra^\circ)}\circ\sfP_0\circ\sfD_\rint\circ\sfE\big)\big|_{\sfT^i(\fra^\circ)}\ =\ \sum_{\ell=0}^\infty\hbar^\ell\sfm_{i,\ell}^\circ
            \end{equation}
        \end{subequations}
        and with $\sfE$ satisfying again the recursion relation~\eqref{eq:looplevelE2}. The interaction vertices $\sfm_i$ in $\sfD_\rint$, as given in~\eqref{eq:YM_higher_products}, lead to products of the colour parts and kinematic functions. Given (composite) fields $\Phi_i=\phi_iX_i\in\fra_1$, we can define colour-stripped interactions $m_i$ by
        \begin{equation}\label{eq:colourStripped}
            \sfm_i(\Phi_1,\ldots,\Phi_i)\ =:\ m_i(\phi_1,\ldots,\phi_i)X_1\cdots X_i
        \end{equation}
        and $\sfD_\rint$ acts on tensor products as in~\eqref{eq:extendedProducts}, e.g.
        \begin{equation}\label{eq:actionMYM}
            \begin{aligned}
                \sfD_\rint(\Phi_1\otimes\Phi_2\otimes\Phi_3)\ &=\ m_2(\phi_1,\phi_2)X_1X_2\otimes\phi_3X_3+\phi_1X_1\otimes m_2(\phi_2,\phi_3)X_2X_3~+\\
                &\kern1cm+m_3(\phi_1,\phi_2,\phi_3)X_1X_2X_3~.
            \end{aligned}
        \end{equation}
        Moreover, $\Delta^*$ acts similarly as in~\eqref{eq:actionDualBVLScalar} on the components $\phi_i$ of $\Phi_i$ by inserting in all possible places of the tensor product of the $\Phi_i$s a complete pair of field and antifield components, 
        \begin{equation}
            \Psi^+_\Theta\ =\ \psi^+_\theta(k,\varepsilon)|a)(b|\eand
            \Psi^\Theta\ =\ \psi^\theta(k,\varepsilon)|b)(a|~,
        \end{equation}
        where $|a)(b|$ is the $(N\times N)$-matrix with the only non-vanishing entry $1$ at position $(a,b)$ and $\Theta$ are multi-indices including particle species (labelled by $\theta$), momenta (labelled by $k$), polarisations (labelled by $\varepsilon$), and colours (labelled by $a$ and $b$). Contractions of $\Theta$ thus imply sums and integrals. 
        
        If $\Delta^*$ is applied once in the recursion, the colour factor of the amplitude contains terms of the form 
        \begin{subequations}
            \begin{equation}\label{eq:actionDualBVLYM1}
                \begin{aligned}
                    &\sum_{a,b=1}^N X_1\otimes\cdots\otimes X_j\otimes|a)(b|\otimes|b)(a|\otimes X_{j+1}\otimes\cdots\otimes X_i\\
                \end{aligned}
            \end{equation}
            and
            \begin{equation}\label{eq:actionDualBVLYM2}
                \begin{gathered}
                    \sum_{a,b=1}^N X_1\otimes\cdots\otimes X_j\otimes|a)(b|\otimes X_{j+1}\otimes\cdots\otimes\,X_k\otimes|b)(a|\otimes X_{k+1}\otimes\cdots\otimes X_i~.
                \end{gathered}
            \end{equation}
        \end{subequations}
        
        Contributing to the amplitude~\eqref{eq:amplitude_YM} are exactly those expressions in which all the tensor products in the colour factors have been turned into matrix products by the $\sfD_{\rint}$. The terms~\eqref{eq:actionDualBVLYM1}, with neighbouring insertion points, enter into planar Feynman diagrams and they come with an additional factor of $N$. The terms~\eqref{eq:actionDualBVLYM2} enter into non-planar Feynman diagrams.
        
        More generally, it is clear that the $\ell$-loop $n$-point amplitude has maximally $t={\rm max}\{\ell,n\}$ traces in its colour factor and that contributions with $t$ traces come with a factor $N^{\ell-t+1}$. Thus, as well-known, planar Feynman diagrams dominate in the large-$N$ limit.
        
        \subsection{One-loop structure}
        
        Let us look at the structure of one-loop scattering amplitudes in more detail. Upon iterating~\eqref{eq:looplevelE2}, we find
        \begin{equation}\label{eq:minimalProductYM2}
            \begin{gathered}
                \sfm_{i,1}^\circ\ =\ \big({\rm pr}_{\sfT^1(\fra^\circ)}\circ\sfP|_{\caO(\hbar^0)}\circ(-\rmi\Delta^*)\circ\sfE|_{\caO(\hbar^0)}\big)\big|_{\sfT^i(\fra^\circ)}~,\\
                \sfP|_{\caO(\hbar^0)}\ =\ \sfP_0\circ(1+\sfD_\rint\circ\sfH_0)^{-1}~,\\
                \sfE|_{\caO(\hbar^0)}\ =\ (1+\sfH_0\circ\sfD_\rint)^{-1}\circ\sfE_0~;
            \end{gathered}
        \end{equation}
        see also~\eqref{eq:hpl_relations}. The form of the interaction vertices and our above considerations directly yield
        \begin{equation}\label{eq:minimalProductYM3}
            \begin{aligned}
                \sfm_{i,1}^\circ(A_1,\ldots,A_i)
                \ &=\ \kappa^{i-1}\Big[N J_{i,1}(1,\ldots,i)\,\rme^{\rmi k_{1i}\cdot x} X_1\cdots X_i~+\\
                &\kern1.5cm+\sum_{j=1}^{i-1}K^j_{i,1}(1,\ldots,i)\,\rme^{\rmi k_{1i}\cdot x} X_1\cdots X_j\,\tr(X_{j+1}\cdots X_i)\Big]\Big|_{k_{1i}^2=0}\\
            \end{aligned}
        \end{equation}
        with $k_{ij}\coloneqq k_i+\cdots+k_j$ for $i\leq j$. The currents $J_{i,1},K^j_{i,1}\in\Omega^1$ contain all the kinematical information and eventually form the one-loop generalisation of the tree-level Berends--Giele current~\cite{Berends:1987me} after symmetrisation.
        
        The general form of the one-loop amplitude thus is
        \begin{equation}\label{eq:YM1loopBern}
            \begin{aligned}
                \caA_{n,1}(A_1,\ldots,A_n)
                \ &=\ N\!\!\sum_{\sigma\in S_n/\IZ_n}\alpha_{n,1}^0(\sigma(1),\ldots,\sigma(n))\,\tr(X_{\sigma(1)}\cdots X_{\sigma(n)})~+\\
                &\kern.5cm+\sum_{m=1}^{n-1}\sum_{\sigma\in S_n/(\IZ_m\times\IZ_{n-m})}\alpha_{n,1}^m(\sigma(1),\ldots,\sigma(n))~\times\\
                &\kern1.5cm\times\tr(X_{\sigma(1)}\cdots X_{\sigma(m)})\tr(X_{\sigma(m+1)}\cdots X_{\sigma(n)})~,
            \end{aligned}
        \end{equation}
        where $\alpha_{n,1}^0$ is a linear combination of (the components of) $J_{n-1,1}$ and the $\alpha_{n,1}^m$ of $K_{n-1,1}^{m-1}$. The result~\eqref{eq:YM1loopBern} was first derived in~\cite{Bern:1990ux} using different methods.   
        
        In~\cite{Bern:1994zx} it was shown that the $\alpha_{n,1}^m$ are linear combinations of the $\alpha_{n,1}^0$ so that the full scattering amplitude can be constructed from its planar part. Explicitly, 
        \begin{equation}\label{eq:rel_planar_non_planar}
            \alpha_{n,1}^m(1,\ldots,n)\ =\ (-1)^{m}\!\!\!\sum_{\sigma\in{\rm COP}_{m,n}}\alpha_{n,1}^0(\sigma(1),\ldots,\sigma(n))~,
        \end{equation}
        where ${\rm COP}_{m,n}$ are all permutations of $(1,\ldots,n)$ which preserve the position of $n$ as well as the cyclic orders of $(1,\ldots,m)$ and $(m+1,\ldots,n)$.
        
        The relation~\eqref{eq:rel_planar_non_planar} can be derived from our recursion relation, but the derivation simplifies significantly if we use the strictification theorem for homotopy algebras (see e.g.~\cite{Berger:0512576}): any $A_\infty$-algebra is quasi-isomorphic (read: equivalent for all physical purposes, cf.~\cite{Macrelli:2019afx,Jurco:2018sby}) to a strict $A_\infty$-algebra, which is an $A_\infty$-algebra with $\sfm_i=0$ for $i\geq 3$. YM theory admits a first-order formulation which constitutes a strictification, see~\cite{Costello:2006ey,Costello:2007ei,Costello:2011aa,Rocek:2017xsj,Macrelli:2019afx,Jurco:2018sby} (see also~\cite{Okubo:1979gt,Witten:2003nn}) for the $L_\infty$-algebra description and the quasi-isomorphism, and we readily apply our formalism. Specifically, we compute again scattering amplitudes using formulas~\eqref{eq:computing_amplitudes}, but now $\sfm_3=0$, which simplifies the discussion, and the plane waves have to be replaced by their pre-image under the (strict!) isomorphism that links the minimal models of the original $A_\infty$-algebra and of its minimal model.
        
        As in the ordinary case, $m_2$ is anti-symmetric also in the strict case. Moreover, $\sfm^\circ_2$ cannot change the order of the colour parts $X_i$, and so, $\alpha^m_{n,1}$ arises from the terms
        \begin{equation}\label{eq:non-planar_trees}
            \begin{aligned}
                \sum_{k=m}^{n-1}\sum_{\sigma\in C_m}&\Big\langle\sfe(A_{n}),\caM\big(\sfD_\rtree(A_{m+1}\otimes\cdots\otimes A_k\otimes\sfh(\Psi^+_\Theta)\,\otimes\\
                &~~~~~\otimes A_{\sigma(1)}\otimes\cdots\otimes A_{\sigma(m)}\otimes\Psi^\Theta\otimes A_{k+1}\otimes\cdots\otimes A_{n-1})\big)~+\\
                &+\caM\big(\sfD_\rtree(A_{m+1}\otimes\cdots\otimes A_k\otimes\Psi^\Theta\otimes A_{\sigma(1)}\,\otimes\\
                &~~~~~\otimes\cdots\otimes A_{\sigma(m)}\otimes\sfh(\Psi^+_\Theta)\otimes A_{k+1}\otimes\cdots\otimes A_{n-1})\big)
                \Big\rangle~,
            \end{aligned}
        \end{equation}
        where $\sfD_\rtree\coloneqq\sfD_\rint\circ(\sfH\circ\sfD_\rint)^{n-1}$ produces a formal sum of full binary trees with $n+1$ leaves corresponding to the $n+1$ arguments and nodes corresponding to the map $\sfm_2$ applied to their children. We call these trees non-planar trees and the leaves corresponding to the $A_{1},\ldots,A_{m}$ inner leaves, while all other leaves are outer leaves. For any tree, the sequence of arguments corresponding to the leaves of the tree will be called its leaf sequence. 
        
        Similarly, the planar trees relevant in the planar contributions arise from expressions 
        \begin{equation}\label{eq:planar_trees}
            \begin{aligned}
                &\sum_{k=0}^{n-1}\sum_{\sigma\in {\rm COP}_{m,n}}\big\langle\sfe(A_{n}),\caM(\sfD_\rtree(A_{\sigma(1)}\otimes\cdots\otimes A_{\sigma(k)}\,\otimes\\[-0.3cm]
                &\kern3.5cm\otimes(\sfh(\Psi^+_\Theta)\otimes\Psi^\Theta+\Psi^\Theta\otimes\sfh(\Psi^+_\Theta))\otimes A_{\sigma(k+1)}\otimes\cdots\otimes A_{\sigma(n-1)}))\big\rangle~.
            \end{aligned}
        \end{equation}
        For both the non-planar and planar trees, the linear function $\caM$ assigns a combinatorial factor to each tree, arising from the various sequences of the operations $\sfH\circ\sfD_\rint$ and $\sfH\circ\Delta^*$ in the recursion relation~\eqref{eq:looplevelE2}.
        
        Upon stripping off the colour factor in each tree, $\tr(X_1\cdots X_m)\tr(X_{m+1}\cdots X_n)$, we obtain two formal sums of binary trees with nodes corresponding to $m_2$ and leaf sequences consisting of $a_i$, $\psi^\theta(k,\varepsilon)$ and $\sfh(\psi_\theta^+(k,\varepsilon))$.
        
        There is now a one-to-one correspondence between the two sets of full binary trees with leaf sequence $A_1,\ldots,A_k$ and with leaf sequence $A_k,\ldots,A_1$, by inverting the order of children in each of the $k-1$ nodes (`flipping the nodes'), which gives rise to a factor of $(-1)^{k-1}$.
        
        In each non-planar binary tree with inner leaves, we can now flip common ancestor of a $\psi$, turning inner leaves into outer leaves. We start from common ancestors closest to the leaves. In each flip, $k$ inner leaves are turned into outer leaves, and together with the initial flip, fully reversing their ordering leads to a relative factor of $(-1)^k$. We stop this process when all $m$ inner leaves have become outer leaves, with a relative factor of $(-1)^m$.
        
        This map from non-planar to planar trees is clearly invertible. It is, however, not surjective since its image does not contain planar trees which have vertices who have a $\psi$ and a root of a subtree containing both inner and outer leaves as descendants. These, however, cancel pairwise: pick any outer leaf, and flip the first common ancestor with an inner leaf. This leads to a negative contribution from another tree, which is included in~\eqref{eq:planar_trees} due to the sum over the COP permutations.
        
        It remains to compare the multiplicities $\caM$ for non-planar and planar trees. Flipping a node does not change the combinatorial factor for applying $\sfH\circ\sfD_\rint$ in different ways. It can, however, affect the multiplicity arising from applying $\sfH\circ\Delta^*$ at different positions since in the planar trees, inner and outer leaves can be joined to subtrees before applying $\sfH\circ\Delta^*$, which was not possible in the non-planar case. These subtrees are of the type discussed in the previous paragraph and they cancel again pairwise.
        
        \section{Conclusions}\label{sec:conclusions}
        
        We showed that full quantum scattering amplitudes of quantum field theories can be conveniently described in terms of minimal models of cyclic quantum $A_\infty$-algebras. This description allows for recursion relations for currents, which reproduce and generalise known recursion relations. As an application, we re-derived known results for one-loop YM scattering amplitudes using our formalism. We conclude that the homotopy algebraic perspective is very useful for understanding the structure of scattering amplitudes. 

        In our discussion, we made use of $A_\infty$-algebras as they turned out to be more suitable from the point of view of stripping off colour as done e.g.~in~\eqref{eq:colourStripped} and also in view of discussing the large-$N$ or planar limit. Since the BV formalism naturally produces an $L_\infty$-algebra, one may wonder whether the transition to $A_\infty$-algebras involves some ambiguity. In general, this would be the case, but for field theories this is usually fixed as we shall explain now.
        
        The higher products of the $L_\infty$-algebra for scalar theory can be naturally identified with the higher products of an $A_\infty$-algebra. In particular, they agree with their own graded antisymmetrisation and a unique, preferred choice of $A_\infty$-algebra is given by the $L_\infty$-algebra. 
        
        For Yang--Mills theory, the colour stripping involves a unique factorisation of the $L_\infty$-algebra $\frl$ of Yang--Mills theory as the tensor product 
        \begin{equation}
            \frl\ =\ \frg\otimes \frc~,
        \end{equation}
        where $\frg$ is the gauge Lie algebra and $\frc$ is a `colour-stripped' homotopy algebra encoding the kinematics, which is a specialisation of an $A_\infty$-algebra known as a strong homotopy commutative algebra or $C_\infty$-algebra. If $\frg$ is a matrix Lie algebra, then we have the unique $A_\infty$-algebra
        \begin{equation}
            \fra\ =\ \frg\otimes \frc
        \end{equation}
        describing Yang--Mills theory, where $\frg$ is now regarded as an associative (matrix) algebra and $\frc$ is again the kinematical $C_\infty$-algebra. Note, however, that $\fra$ itself is not a $C_\infty$-algebra. 
        
        Thus, if we impose the condition that the generalisation from $\frl$ to $\fra$ is compatible with the factorisation in a colour-stripped $C_\infty$-algebra, we obtain a unique generalisation to $A_\infty$-algebras. We shall report on the full details in the paper~\cite{Borsten:2020XXX}, where we will explain the homotopy algebra structures we used in much greater detail and apply our formalism to a number of other problems. 
        
        In future work, we also plan to address the peculiarities related to the singular nature of the BV Laplacian for infinite-dimensional function spaces, where one needs to modify the quantum master equation into a renormalised quantum master equation as done in~\cite{Costello:2007ei}.
            
    \end{body}
        
\end{document}